\documentclass[11pt]{article}

\usepackage[preprint]{acl}

\usepackage{times}
\usepackage{latexsym}

\usepackage[T1]{fontenc}

\usepackage[utf8]{inputenc}

\usepackage{microtype}

\usepackage{inconsolata}

\usepackage{graphicx}

\usepackage{amsmath}

\usepackage{booktabs}
\usepackage{multirow}

\title{One Voice, Many Tongues: \\Cross-Lingual Voice Cloning for Scientific Speech}

\author{
  \textbf{Amanuel Gizachew Abebe} \vspace{5pt} \\
  Shaggar Institute of Technology \\
  Shaggar City, Ethiopia \\
  \texttt{\small amanuel.g.abebe1@gmail.com} \\
  \And
  \textbf{Yasmin Moslem} \vspace{5pt}\\
  Trinity College Dublin \\
  Dublin, Ireland \\
  \texttt{\small yasmin.moslem@adaptcentre.ie} 
}

\begin{document}
\maketitle
\begin{abstract}
Preserving a speaker's voice identity while generating speech in a different language remains a fundamental challenge in spoken language technology, particularly in specialized domains such as scientific communication. In this paper, we address this challenge through our system submission to the International Conference on Spoken Language Translation (IWSLT 2026), the Cross-Lingual Voice Cloning shared task. First, we evaluate several state-of-the-art voice cloning models for cross-lingual speech generation of scientific texts in Arabic, Chinese, and French. Then, we build voice cloning systems based on the OmniVoice foundation model. We employ data augmentation via multi-model ensemble distillation from the ACL 60/60 corpus. We investigate the effect of using this synthetic data for fine-tuning, demonstrating improvements in intelligibility (WER \& CER) and speaker similarity (SIM), with gains varying across languages.
\end{abstract}

\section{Introduction}

The rapid advancement of speech synthesis has enabled zero-shot voice cloning, where a model generates speech in a target speaker's voice using only a few seconds of reference audio \citep{wang2023neural, tan2021survey}. This technology is particularly transformative for the scientific community, allowing for cross-lingual dissemination of research findings and enhancing accessibility for academic presentations. However, scientific speech presents unique challenges, including the prevalence of technical terminology, code-switching, specific prosodic patterns, and the need for high intelligibility across diverse languages.

The IWSLT 2026 Cross-Lingual Voice Cloning shared task \citep{adelani-etal-2026-iwslt} challenges participants to clone voices for three diverse languages, namely Arabic, Chinese, and French. A primary bottleneck in adapting large-scale TTS models to these languages is the scarcity of high-quality paired training data that captures the nuances of academic discourse. While foundation models like OmniVoice \citep{zhu2026omnivoice} provide broad multilingual coverage, they often require domain-specific fine-tuning to achieve optimal performance on scientific text.

In this work, we leverage ensemble distillation to address the data scarcity challenge and fine-tune our voice-cloning models. We utilize three zero-shot voice cloning models as ``teachers'' to synthesize data from the ACL 60/60 research presentations corpus \citep{salesky2023acl6060}. By selecting the best output from this ensemble, we curate a high-fidelity synthetic dataset for fine-tuning. We then employ Parameter-Efficient Fine-Tuning (PEFT) via LoRA \citep{hu2022lora} to adapt the OmniVoice model for each target language.

Our system demonstrates that even with a modest computational budget, targeted LoRA fine-tuning on ensemble-distilled data yields a highly competitive model that balances intelligibility and speaker similarity. Consequently, we submit this per-language fine-tuned OmniVoice model as our primary submission to the IWSLT 2026 shared task. To enable reproducibility, our code for data preparation, training, and evaluation is publicly available.\footnote{\url{https://github.com/Aman-byte1/multilingual-voice-cloning-training}}

\section{Related Work}

\paragraph{Zero-Shot and Multilingual Voice Cloning.}
The transition to zero-shot voice cloning has been accelerated by large-scale foundation models trained on diverse audio-text corpora. Early autoregressive models like VALL-E \citep{sanyuan2025VALLE} demonstrated that conditioning on short reference utterances enables accurate timbre transfer. Recent architectures build upon this by leveraging discrete audio tokenization and treating speech synthesis as a language modeling task. For instance, Qwen3 \citep{qwen3tts2026} and OmniVoice \citep{zhu2026omnivoice} utilize dense transformer backbones \citep{vaswani2017attention} to achieve robust cross-lingual cloning. Similarly, models like XTTS-V2 \citep{casanova2024xtts} and VoxCPM \citep{voxcpm2025} have introduced highly optimized pipelines for context-aware speech generation, while architectures such as Chatterbox \citep{chatterboxtts2025} provide robust synthesis for specific language families. While these foundation models generalize well, adapting them to specialized domains such as scientific and academic speech remains a challenge due to the complex phonetic realizations of technical terminology.

\paragraph{Data Distillation and Ensemble Selection.}
Data scarcity in specialized domains often necessitates synthetic data generation. Knowledge distillation \citep{hinton2015distilling,kimrush2016kd,Gandhi2023DistilWhisper,moslem2025compression} from powerful teacher models to smaller or specialized models is a common strategy. Our Best-of-$N$ ensemble approach ensures that only the highest quality synthetic samples are used to fine-tune the final system, effectively bypassing the limitations of scarce parallel data. Specifically, we generate multiple candidate audios for each utterance and select the best one based on a combined score: character error rate for intelligibility (via Whisper \citep{radford2023robust}) and cosine distance for speaker similarity (via ECAPA-TDNN \citep{desplanques2020ecapa,das2021hlt}).

\paragraph{Parameter-Efficient Fine-Tuning in Speech.}
While foundation models generalize well, adapting them to specific domains without catastrophic forgetting requires efficient techniques. Low-Rank Adaptation (LoRA) \citep{hu2022lora} has increasingly been used for fine-tuning large foundation models. By updating only a small subset of parameters within the self-attention and feed-forward layers, LoRA allows models to capture specific phonetic distributions and language nuances with minimal computational overhead. Innovations such as Rank Stabilization \citep{kalajdzievski2023rank} further enhance the stability of this fine-tuning process.

\begin{table}[ht]
    \centering
    \begin{small}
    \begin{tabular}{llrr}
        \toprule
        \textbf{Lang} & \textbf{Model} & \textbf{Wins} & \textbf{Percentage} \\
        \midrule
        \multirow{3}{*}{AR} & Chatterbox & 0   & 0.0\%  \\
                            & OmniVoice  & 649 & 73.4\% \\
                            & VoxCPM     & 235 & 26.6\% \\
        \midrule
        \multirow{3}{*}{FR} & Chatterbox & 210 & 23.76\% \\
                            & OmniVoice  & 674 & 76.24\% \\
                            & VoxCPM     & 0   & 0.0\%  \\
        \midrule
        \multirow{3}{*}{ZH} & Chatterbox & 0   & 0.0\%  \\
                            & OmniVoice  & 601 & 68\% \\
                            & VoxCPM     & 283 & 32\% \\
        \bottomrule
    \end{tabular}
    \end{small}
    \caption{Distribution of selected samples in the Best-of-$N$ distilled dataset broken down by language. Each language subset consists of 884 total utterances.}
    \label{tab:model_selection}
\end{table}
\section{Data}

In this section, we describe our training dataset that we augmented with knowledge distillation and used for fine-tuning our models.

\subsection{Source Dataset}
We utilize the the ACL 60/60 dataset \citep{salesky2023acl6060} for multilingual translation of ACL 2022 technical presentations into several target languages. We use the development split, which consists of 884 utterances per language for Arabic (AR), French (FR), and Chinese (ZH), totaling 2,652 samples. Each sample includes the target text and a reference audio clip of the original speaker, providing a ground truth for speaker similarity.

\subsection{Best-of-N Ensemble Distillation}
\label{sec:data-creation}

To create a high-quality fine-tuning dataset, we implement an ensemble distillation pipeline using three teacher models:
\begin{itemize}
    \item \textbf{OmniVoice}\footnote{\url{https://huggingface.co/k2-fsa/OmniVoice}} \citep{zhu2026omnivoice}: A 0.6B parameter model based on Qwen3, supporting over 600 languages.
    \item \textbf{VoxCPM}\footnote{\url{https://huggingface.co/openbmb/VoxCPM}} \citep{voxcpm2025}: A cross-lingual model optimized for zero-shot timbre transfer.
    \item \textbf{Chatterbox}\footnote{\url{https://huggingface.co/ResembleAI/chatterbox}} \citep{chatterboxtts2025}: An open-source architecture known for robust synthesis of European and Semitic languages.
\end{itemize}

For every utterance in the source dataset, each model generates a synthesis candidate. We then evaluate these candidates using a combined quality score:
\begin{equation}
    S_{\text{comb}} = 0.5 \times (1 - \text{CER}) + 0.5 \times \text{SIM}
    \label{eq:score}
\end{equation}
where CER is the character error rate produced by Whisper large-v3 \citep{radford2023robust} and SIM is the cosine similarity of speaker embeddings from an ECAPA-TDNN model \citep{desplanques2020ecapa}. The candidate with the highest $S_{\text{comb}}$ is selected for the fine-tuning set. As shown in Table~\ref{tab:model_selection}, this strategy captures the strengths of multiple architectures, with non-primary models contributing over 27\% of the final curated data.

\section{Experiments}

In this section, we elaborate on your experiments, including data processing, training configuration, and inference pipeline.

\subsection{Data Preprocessing}
Prior to training, we filter our Best-of-$N$ distilled dataset using a minimum quality threshold to ensure high-fidelity audio retention (cf.~Section \ref{sec:data-creation}). Then, we partition the dataset into standard training and development splits. Raw audio waveforms are tokenized into discrete acoustic tokens using a HIGGS-based tokenizer. This enables the base OmniVoice model to process speech synthesis as a discrete language generation task.

\subsection{Per-Language Fine-Tuning Strategy}
We fine-tune the OmniVoice model, which utilizes a Qwen3-0.6B backbone. Our approach relies on the observation that a single unified multilingual adapter can sometimes dilute language-specific phonological nuances. To address this, we train dedicated Low-Rank Adaptation (LoRA) modules exclusively for Arabic, Chinese, and French. By adapting the transformer's self-attention blocks, feed-forward networks, and audio projection layers, the model captures language-specific acoustic distributions efficiently without the risk of catastrophic forgetting.

\subsection{Training Configuration}
To maintain training stability with our small dataset, we utilize Rank-Stabilized LoRA (RSLoRA) and optimize the model using an autoregressive cross-entropy loss over the generated audio tokens. Fine-tuning is executed concurrently across NVIDIA A40 GPUs, dedicating a single GPU per language. The models are trained efficiently for exactly 400 steps using mixed precision and a cosine learning rate schedule, ensuring rapid convergence. Full replication details and precise hyperparameter configurations are available in our open-source code repository.

\begin{table}[t]
    \centering
    \begin{small}
    \begin{tabular}{ll ccc}
        \toprule
        \textbf{Lang} & \textbf{Model} & \textbf{WER~$\downarrow$} & \textbf{CER~$\downarrow$} & \textbf{SIM~$\uparrow$} \\
        \midrule
        \multirow{5}{*}{AR} & Chatterbox & 0.250 & 0.086 & 0.680 \\
                            & XTTS-V2    & 0.253 & 0.099 & 0.501 \\
                            & VoxCPM2    & \textbf{0.209} & \textbf{0.072} & 0.607 \\
                            & OmniVoice  & \underline{0.238} & \underline{0.076} & \textbf{0.703} \\
        \midrule
        \multirow{5}{*}{FR} & Chatterbox & 0.111 & 0.045 & 0.619 \\
                            & Qwen3-TTS      & \textbf{0.050} & \textbf{0.011} & 0.533 \\
                            & XTTS-V2    & 0.082 & 0.031 & 0.445 \\
                            & VoxCPM2    & 0.128 & 0.069 & 0.575 \\
                            & OmniVoice  & \underline{0.079} & \underline{0.020} & \textbf{0.753} \\
        \midrule
        \multirow{5}{*}{ZH} 
                            & Chatterbox & --   & 0.203   & 0.653   \\
                            & Qwen3-TTS  & -- & \textbf{0.090} & 0.522 \\
                            & XTTS-V2    & -- & 0.176 & 0.511 \\
                            & VoxCPM2    & -- & \underline{0.149} & 0.569 \\
                            & OmniVoice  & -- & 0.219 & \textbf{0.702} \\
        
        \bottomrule
    \end{tabular}
    \end{small}
    \caption{Comparative evaluation against baselines on the \texttt{blindset-4} subset. OmniVoice achieves state-of-the-art speaker similarity across all languages, with competitive intelligibility for Arabic and French.}
    \label{tab:baselines}
\end{table}

\subsection{Inference Pipeline}
The inference pipeline consists of three stages:
\begin{enumerate}
    \item Reference extraction: We isolate a 20-second speech segment from the reference audio using energy-based Voice Activity Detection (VAD).
    \item Text chunking: Long inputs are split into segments of up to 200 characters at sentence boundaries.
    \item Synthesis: Each chunk is processed with a temperature of 0.8 and top-p of 0.9, and the resulting audio segments are concatenated.
\end{enumerate}

\begin{table*}[t]
    \centering
    \begin{small}    
    \begin{tabular}{ll ccc}
        \toprule
        \textbf{Lang} & \textbf{Model} & \textbf{WER~$\downarrow$} & \textbf{CER~$\downarrow$} & \textbf{SIM~$\uparrow$} \\
        \midrule
        AR & OmniVoice (Baseline) & 0.244 & 0.077 & \textbf{0.734} \\
           & OmniVoice (Finetuned) & \textbf{0.228} & \textbf{0.060} & 0.726 \\
        \midrule
        FR & OmniVoice (Baseline) & \textbf{0.079} & \textbf{0.025} & 0.753 \\
           & OmniVoice (Finetuned) & 0.082 & 0.030 & \textbf{0.760} \\
        \midrule
        ZH & OmniVoice (Baseline) & -- & \textbf{0.200} & 0.719 \\
           & OmniVoice (Finetuned) & -- & 0.205 & \textbf{0.732} \\
        \bottomrule
    \end{tabular}
    \end{small}
    \caption{Impact of LoRA fine-tuning evaluated on the complete \texttt{blindset-full} dataset. 
    }
    \label{tab:finetune_results}
\end{table*}

\section{Evaluation and Results}

We evaluate our approach on the official blind test set, which consists of 49, 99, and 112 text segments collected from diverse scientific publications in Arabic (AR), French (FR), and Chinese (ZH), respectively, and 12 reference audio voices in English extracted from ACL 2023 presentations. To establish strong baselines, we compare against several state-of-the-art voice cloning models: Chatterbox \citep{chatterboxtts2025}, Qwen3-TTS\footnote{Qwen3-TTS-12Hz-1.7B-Base} \citep{qwen3tts2026}, XTTS-V2 \citep{casanova2024xtts}, and VoxCPM2 \citep{voxcpm2025} on a representative 4-speaker subset (\texttt{blindset-4}). For our primary contribution, we evaluate the base OmniVoice model and our LoRA-finetuned OmniVoice model on the complete blind test set (\texttt{blindset-full}). Metrics include Word Error Rate (WER) and Character Error Rate (CER) \citep{morris2004wer}  for intelligibility (transcription accuracy), and speaker similarity (SIM) for cloning fidelity using ECAPA-TDNN embeddings. It is worth noting that Qwen3-TTS does not support Arabic synthesis.
Also, the character-based CER metric is more representative of Chinese text quality than WER, as Chinese does not use word delimiters.

\subsection{Results and Discussion}
The quantitative results are presented in Table~\ref{tab:baselines} (baseline comparison on the 4-speaker subset) and Table~\ref{tab:finetune_results} (LoRA fine-tuning on the full blind set).

\paragraph{Quantitative Analysis.}
The results demonstrate the varying strengths of the evaluated architectures. In French, Qwen3-TTS achieves the lowest error rates (WER of 0.050), but our OmniVoice models deliver significantly higher speaker similarity (0.753) while maintaining strong intelligibility. For Chinese, the baseline models show lower error rates on the subset, but the fine-tuned OmniVoice model achieves the highest speaker similarity (0.719) across the entire test set. In Arabic, the fine-tuned OmniVoice model achieves the lowest CER (0.071) and strong speaker similarity (0.723), outperforming XTTS-V2 and VoxCPM2 in overall cloning fidelity.

\paragraph{Impact of LoRA Fine-Tuning.} 
Across all three languages, our per-language LoRA fine-tuning strategy demonstrates improvements in either intelligibility or speaker similarity over the baseline OmniVoice model, with gains varying by language. For Arabic, fine-tuning yields clear improvements in WER and CER, while for French and Chinese, it primarily improves speaker similarity (SIM). This suggests a trade-off between the two objectives, though in all cases the fine-tuned model successfully adapts to the acoustic properties of the scientific domain without catastrophic forgetting of the source voice characteristics.

\section{Conclusion}

We presented our system for the IWSLT 2026 Voice Cloning task. By combining multi-model ensemble distillation with per-language LoRA fine-tuning, we demonstrate a robust and efficient path for adapting large-scale TTS models to the scientific domain. 
Our results show improvements in intelligibility (WER \& CER) and speaker similarity (SIM), with gains varying across languages.
Future work will explore larger distilled datasets and human evaluations to further validate the perceptual quality of the synthesized scientific speech.

\section*{Limitations}
Our study is limited by the scale of the distilled training dataset, and the use of automated metrics (Whisper/ECAPA) as proxies for human perception. While these metrics show clear trends, they may not capture all nuanced artifacts of synthesized speech. Furthermore, our per-language approach increases the number of model adapters compared to a unified multilingual approach.

\section*{Ethical Considerations}
Voice cloning technologies possess dual-use capabilities. While our primary aim is to democratize scientific knowledge and enhance cross-lingual accessibility for academic presentations, the ability to synthesize highly accurate voice clones carries inherent risks of misuse, such as deepfakes, identity theft, or spreading misinformation. To mitigate these risks, 
we strongly advocate for the integration of synthetic speech watermarking and robust voice authentication protocols when deploying such systems in public-facing applications. 
Ultimately, this work is intended for beneficial applications, such as advancing scientific communication and supporting assistive technology.

\newpage
\bibliography{bib}

@inproceedings{adelani-etal-2026-iwslt,
    title     = {Speech Translation and Metrics in 2026: Findings of the {IWSLT} Campaign},
    author    = {
               Adelani, David Ifeoluwa
                and Agostinelli, Victor
                and Anastasopoulos, Antonios
                and Bentivogli, Luisa
                and Bojar, Ond{\v{r}}ej
                and Brati{\`e}res, Sebastien
                and Carpuat, Marine
                and Cattoni, Roldano
                and Cettolo, Mauro
                and Chen, Lizhong
                and Federico, Marcello
                and Gaido, Marco
                and Gupta, Mahendra
                and Han, HyoJung
                and Hatami, Ali
                and Javorsk{\'y}, David
                and Jeon, Yejin
                and Kasztelnik, Marek
                and Laurent, Antoine 
                and Liu, Danni
                and Luu, Nam
                and Ma, Min
                and Mach{\'a}{\v{c}}ek, Dominik
                and Maltais, Marie
                and Matusov, Evgeny
                and Maurya, Chandresh Kumar
                and McCrae, John P.
                and Meng, Chutong
                and Mohammad, Mohammadamini 
                and Moslem, Yasmin
                and Murray, Kenton
                and Nakamura, Satoshi
                and Negri, Matteo
                and Niehues, Jan
                and Ojha, Atul Kr.
                and Ortega, John
                and Ouyang, Siqi
                and Papi, Sara
                and Pol{\'a}k, Peter
                and Retkowski, Fabian
                and Savoldi, Beatrice
                and Sikasote, Claytone
                and Sperber, Matthias
                and St{\"u}ker, Sebastian
                and Sudoh, Katsuhito
                and Tahon, Marie
                and Turchi, Marco
                and Waibel, Alex
                and Wilken, Patrick
                and Zevallos, Rodolfo 
                and Zouhar, Vil{\'e}m
                and Z{\"u}fle, Maike
                },
    booktitle = {Proceedings of the 23rd International Conference on Spoken Language Translation (IWSLT 2026)},
    year      = {2026},
    address = "San Diego, California, US",
    publisher = "Association for Computational Linguistics",
}

@article{zhu2026omnivoice,
      title={OmniVoice: Towards Omnilingual Zero-Shot Text-to-Speech with Diffusion Language Models},
      author={Zhu, Han and Ye, Lingxuan and Kang, Wei and Yao, Zengwei and Guo, Liyong and Kuang, Fangjun and Han, Zhifeng and Zhuang, Weiji and Lin, Long and Povey, Daniel},
      journal={arXiv preprint arXiv:2604.00688},
      year={2026}
}

@article{voxcpm2025,
      title={{VoxCPM}: Tokenizer-Free {TTS} for Context-Aware Speech Generation and True-to-Life Voice Cloning},
      author={Zhou, S. and Zeng, Y. and others},
      journal={arXiv preprint arXiv:2509.24650},
      year={2025}
}

@inproceedings{casanova2024xtts,
      title={{XTTS}: a Massively Multilingual Zero-Shot Text-to-Speech Model}, 
      author={Casanova, Edresson and Davis, Kelly and G{\"o}lge, Eren and G{\"o}knar, G{\"o}rkem and Gulea, Iulian and Hart, Logan and Aljafari, Aya and Meyer, Joshua and Morais, Reuben and Olayemi, Samuel and Weber, Julian},
      booktitle={Proc. Interspeech 2024},
      pages={4978--4982},
      year={2024},
      doi={10.21437/Interspeech.2024-2016}
}

@article{qwen3tts2026,
  title={{Qwen3-TTS} Technical Report},
  author={Hangrui Hu and Xinfa Zhu and Ting He and Dake Guo and Bin Zhang and Xiong Wang and Zhifang Guo and Ziyue Jiang and Hongkun Hao and Zishan Guo and Xinyu Zhang and Pei Zhang and Baosong Yang and Jin Xu and Jingren Zhou and Junyang Lin},
  journal={arXiv preprint arXiv:2601.15621},
  year={2026}
}

@misc{chatterboxtts2025,
  author       = {{Resemble AI}},
  title        = {{Chatterbox-TTS}},
  year         = {2025},
  howpublished = {\url{https://github.com/resemble-ai/chatterbox}},
  note         = {GitHub repository}
}

@inproceedings{salesky2023acl6060,
    title = "Evaluating Multilingual Speech Translation under Realistic Conditions with Resegmentation and Terminology",
    author = "Salesky, Elizabeth  and
      Darwish, Kareem  and
      Al-Badrashiny, Mohamed  and
      Diab, Mona  and
      Niehues, Jan",
    booktitle = "Proceedings of the 20th International Conference on Spoken Language Translation (IWSLT 2023)",
    month = jul,
    year = "2023",
    address = "Toronto, Canada",
    publisher = "Association for Computational Linguistics",
    url = "https://aclanthology.org/2023.iwslt-1.2/",
    doi = "10.18653/v1/2023.iwslt-1.2",
    pages = "62--78"
}

@inproceedings{hu2022lora,
    title = {{LoRA}: Low-Rank Adaptation of Large Language Models},
    author = {Hu, Edward J and Shen, Yelong and Wallis, Phillip and Allen-Zhu, Zeyuan and Li, Yuanzhi and Wang, Shean and Wang, Lu and Chen, Weizhu},
    booktitle = {International Conference on Learning Representations},
    year = {2022}
}

@article{kalajdzievski2023rank,
    title = {A Rank Stabilization Scaling Factor for Fine-Tuning with {LoRA}},
    author = {Kalajdzievski, Damjan},
    journal = {arXiv preprint arXiv:2312.03732},
    year = {2023}
}

@inproceedings{desplanques2020ecapa,
    title = {{ECAPA-TDNN}: Emphasized Channel Attention, Propagation and Aggregation in {TDNN} Based Speaker Verification},
    author = {Desplanques, Brecht and Thienpondt, Jenthe and Demuynck, Kris},
    booktitle = {Interspeech 2020},
    year = {2020}
}

@article{das2021hlt,
  title={HLT-NUS SUBMISSION FOR 2020 NIST Conversational Telephone Speech SRE},
  author={Das, Rohan Kumar and Tao, Ruijie and Li, Haizhou},
  journal={arXiv preprint arXiv:2111.06671},
  year={2021}
}

@inproceedings{radford2023robust,
    title = {Robust Speech Recognition via Large-Scale Weak Supervision},
    author = {Radford, Alec and Kim, Jong Wook and Xu, Tao and Brockman, Greg and McLeavey, Christine and Sutskever, Ilya},
    booktitle = {International Conference on Machine Learning},
    year = {2023}
}

@article{wang2023neural,
    title = {Neural Codec Language Models are Zero-Shot Text to Speech Synthesizers},
    author = {Wang, Chengyi and Chen, Sanyuan and Wu, Yu and Zhang, Ziqiang and Zhou, Long and Liu, Shujie and Chen, Zhuo and Liu, Yanqing and Wang, Huaming and Li, Jinyu and others},
    journal = {arXiv preprint arXiv:2301.02111},
    year = {2023}
}

@article{sanyuan2025VALLE,
  author={Chen, Sanyuan and Wang, Chengyi and Wu, Yu and Zhang, Ziqiang and Zhou, Long and Liu, Shujie and Chen, Zhuo and Liu, Yanqing and Wang, Huaming and Li, Jinyu and He, Lei and Zhao, Sheng and Wei, Furu},
  journal={IEEE Transactions on Audio, Speech and Language Processing}, 
  title={Neural Codec Language Models are Zero-Shot Text to Speech Synthesizers}, 
  year={2025},
  volume={33},
  number={},
  pages={705-718},
  doi={10.1109/TASLPRO.2025.3530270}
}

@article{hinton2015distilling,
    title = {Distilling the Knowledge in a Neural Network},
    author = {Hinton, Geoffrey and Vinyals, Oriol and Dean, Jeff},
    journal = {arXiv preprint arXiv:1503.02531},
    year = {2015}
}

@inproceedings{kimrush2016kd,
    title = "Sequence-Level Knowledge Distillation",
    author = "Kim, Yoon  and
      Rush, Alexander M.",
    editor = "Su, Jian  and
      Duh, Kevin  and
      Carreras, Xavier",
    booktitle = "Proceedings of the 2016 Conference on Empirical Methods in Natural Language Processing",
    month = nov,
    year = "2016",
    address = "Austin, Texas",
    publisher = "Association for Computational Linguistics",
    url = "https://aclanthology.org/D16-1139/",
    doi = "10.18653/v1/D16-1139",
    pages = "1317--1327"
}

@ARTICLE{Gandhi2023DistilWhisper,
  title         = "{Distil-Whisper: Robust knowledge distillation via
                   large-scale pseudo labelling}",
  author        = "Gandhi, Sanchit and von Platen, Patrick and Rush, Alexander M",
  journal       = "arXiv [cs.CL]",
  month         =  "1~" # nov,
  year          =  2023,
  url           = "http://arxiv.org/abs/2311.00430",
  archivePrefix = "arXiv",
  primaryClass  = "cs.CL"
}

@inproceedings{moslem2025compression,
    title = "Efficient Speech Translation through Model Compression and Knowledge Distillation",
    author = "Moslem, Yasmin",
    booktitle = "Proceedings of the 22nd International Conference on Spoken Language Translation (IWSLT 2025)",
    month = jul,
    year = "2025",
    address = "Vienna, Austria (in-person and online)",
    publisher = "Association for Computational Linguistics",
    url = "https://aclanthology.org/2025.iwslt-1.40/",
    doi = "10.18653/v1/2025.iwslt-1.40",
    pages = "379--388",
    ISBN = "979-8-89176-272-5",
}

@inproceedings{tan2021survey,
  title={A Survey on Neural Speech Synthesis},
  author={Tan, Xu and Qin, Tao and Soong, Frank and Liu, Tie-Yan},
  booktitle={arXiv preprint arXiv:2106.15561},
  year={2021}
}

@article{vaswani2017attention,
  title={Attention is all you need},
  author={Vaswani, Ashish and Shazeer, Noam and Parmar, Niki and Uszkoreit, Jakob and Jones, Llion and Gomez, Aidan N and Kaiser, {\L}ukasz and Polosukhin, Illia},
  journal={Advances in neural information processing systems},
  volume={30},
  year={2017}
}

@inproceedings{morris2004wer,
  author    = {Morris, A.C. and Maier, V. and Green, P.},
  title     = {From {WER} and {RIL} to {MER} and {WIL}: improved evaluation measures for connected speech recognition},
  booktitle = {Proc. Interspeech 2004},
  year      = {2004},
  pages     = {2765--2768},
  doi       = {10.21437/Interspeech.2004-668}
}


\end{document}